# Hybrid GPS-GSM Localization of Automobile Tracking System

Mohammad A. Al-Khedher

Mechatronics Engineering Department, Al-Balqa Applied University, Amman 11134, Jordan,
E-mail: `moh.alkhedher@fet.edu.jo`

*Abstract*

   *An integrated GPS-GSM system is proposed to track vehicles using Google Earth application. The remote module has a GPS mounted on the moving vehicle to identify its current position, and to be transferred by GSM with other parameters acquired by the automobile's data port as an SMS to a recipient station. The received GPS coordinates are filtered using a Kalman filter to enhance the accuracy of measured position. After data processing, Google Earth application is used to view the current location and status of each vehicle. This goal of this system is to manage fleet, police automobiles distribution and car theft cautions.*

## 1. Introduction

   The ability to accurately detect a vehicle's location and its status is the main goal of automobile trajectory monitoring systems. These systems are implemented using several hybrid techniques that include: wireless communication, geographical positioning and embedded applications.

   The vehicle tracking systems are designed to assist corporations with large number of automobiles and several usage purposes. A Fleet management system can minimize the cost and effort of employees to finish road assignments within a minimal time. Besides, assignments can be scheduled in advanced based on current automobiles location. Therefore, central fleet management is essential to large enterprises to meet the varying requirements of customers and to improve the productivity [1].

## 2. Related work

   Many researchers have proposed the use of cutting edge technologies to serve the target of vehicle tracking. These technologies include: Communication, GPS, GIS, Remote Control, server systems and others.

   The proposed tracking system in this paper is designed to track and monitor automobiles' status that are used by certain party for particular purposes, this system is an integration of several modern embedded and communication technologies [2]-[6]. To provide location and time information anywhere on earth, Global Positioning System (GPS) is commonly used as a space-based global navigation satellite system [2]. The location information provided by GPS systems can be visualized using Google Earth [3].

                                                                              75

In wireless data transporting, Global System of Mobile (GSM) and Short Message Service (SMS) technology is a common feature with all mobile network service providers [4, 5]. Utilization of SMS technology has become popular because it is an inexpensive, convenient and accessible way of transferring and receiving data with high reliability [6].

As shown in figure (1), the proposed system consists of: in-vehicle GPS receiver, GSM modems (stationary and in-vehicle), and embedded controller [7]. The users of this application can monitor the location graphically on Google Earth; they also can view other relevant information of each automobile in the fleet [8, 9].

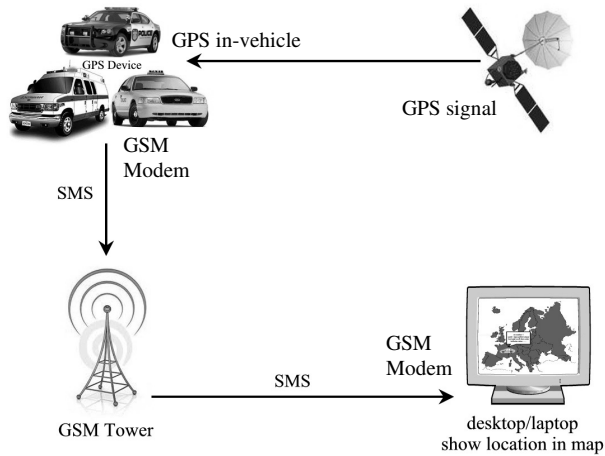

Figure 1. The block diagram of GPS tracking system

The implemented tracking system can be used to monitor various parameters related to safety, emergency services and engine stall [10]. The paper shows an implementation of several modern technologies to achieve a desirable goal of fleet monitoring and management.

## 3. System overview

The system has two main modules, as shown in figure (2). The first module is the tracking device which is attached to the moving automobile. This module composes of: a GPS receiver, Microcontroller and a GSM Modem. The GPS Receiver retrieves the location information from satellites in the form of latitude and longitude real-time readings. The Microcontroller has three main tasks: to read certain engine parameters from automobile data port (OBD-II), to processes the GPS information to extract desired values and to transmit this data to the server using GSM modem by SMS. The chosen engine parameters are: RPM, engine coolant temperature, vehicle speed, percent throttle.

The second module consists of a recipient GSM modem and workstation PC. The modem receives the SMS that includes GPS coordinates and engine parameters. This text is processed using a Visual Basic program to obtain the numeric parameters, which are saved as a Microsoft Office Excel file. The received reading of the GPS is further corrected by Kalman filter. To transfer this information to Google Earth, the Excel file is converted to KML (Keyhole Markup Language) format. Google Earth interprets KML file and shows automobile's location and

International Journal of Computer Science & Information Technology (IJCSIT) Vol 3, No 6, Dec 2011

engine parameters on the map. The system's efficiency is dependable on the sufficiency of the used communication network.

An additional setting could be implemented to interface the system to the car's alarm to alert the owner on his cell phone if the alarm is set off. The automobile's airbag system can also be wired to this system to report severe accidents to immediately alert the police and ambulance service with the location of the accident.

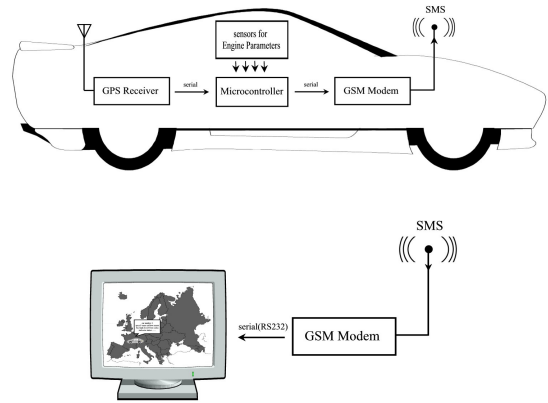

Figure 2. The system architecture: GPS tracking and GSM modules.

## 4. Hardware specification

The tracking unit, as shown in figure (3), consists of two main inputs: The first received input is the GPS output, which has a sentence based on NMEA 0183 standard. The other input is obtained by the automobile data port, typically called ON Board Diagnostics port, version II (OBD-II). The unit sends an SMS using Hayes command (AT Command).

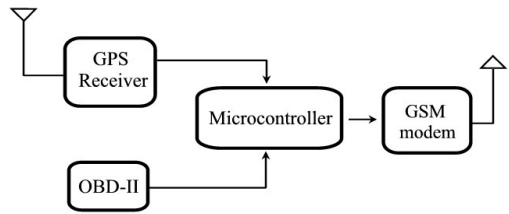

Figure 3. Schematic diagram of in-vehicle tracking unit.

On-Board Diagnostics port (OBD-II) is a universal automotive protocol supported by modern automobiles to retrieve diagnostic errors over a Controller Area Network (CAN) bus of the microcontroller (MCU) [11].The used GSM module is of type SIM900D, this module supports standard AT command and compatible with several GSM networks. Transmission parameters are set to: Baud rate is set at 19200 bps, the data is 8N1 format, and flow control is set to none. For this study, we chose certain parameters to show the status of the engine: RPM, engine coolant temperature, vehicle speed and percent throttle.

The GPS receiver is a MediaTek MT3329. The GPS module supports up to a 10Hz update rate. The microcontroller is the main operational unit of the tracking device. The GPS receiver collects the latitude, longitude and speed information and forwards them to the microcontroller [12]. The GSM module communicates with the microcontroller to send the information package to another GSM Module at the recipient station, all information appears on Google Earth after processing [13]. Figure (4) shows the external view of the tracking unit. The tracking unit is designed to be powered by the automobile battery. However, a power source is built-in the device as an emergency backup.

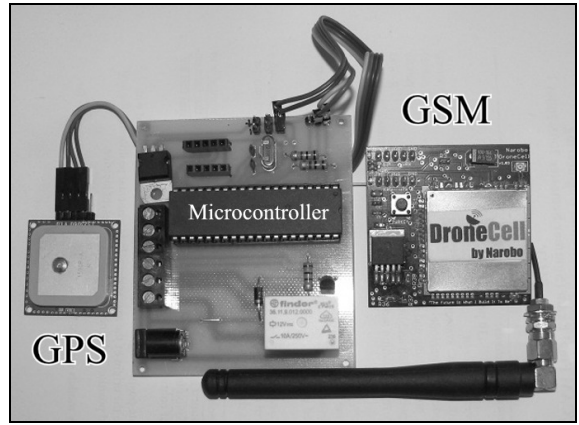

Figure 4. The tracking unit hardware.

## 5. Software specification

In our tracking system we used Google Earth software for tracking and viewing the status of the automobile [14]. Google Earth currently supports most GPS devices. The engaged GPS Module has NMEA 0183 Protocol for transmitting GPS information to a PC. This protocol consists of several sentences, starting with the character $, with a maximum of 79 characters in length. The NMEA Message to read data with both position and time is: $GPRMC [14]. Therefore, only the $GPRMC information is used to determine the location of the automobile to reduce SMS text. The status of the automobile along with $GPRMC information is sent by the GSM modem of type MediaTek MT3329.

Consequently, the recipient GSM, also has NMEA 0183 protocol, receives the transmitted SMS to obtain GPS coordinates and status information of the automobile.

The transmitted GPS data is processed by a Visual Basic program using a Kalman filter to correct the current position. The resulted data of corrected position and automobile parameters is sorted in an Excel sheet. The Excel file is exported to a KML file that is compatible with Google Earth program. Hence, Google Earth will view the location and status of the automobile on the map by reading the KML file. Figure (5) illustrates the block diagram of the recipient module in the system.

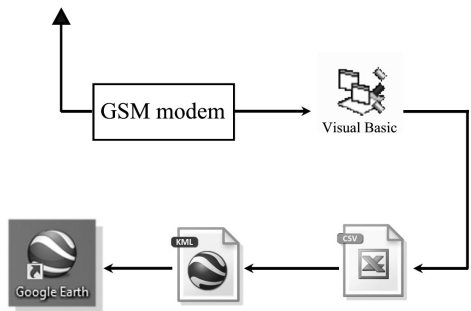

Figure 5. The block diagram of the recipient module in the system.

The KML file, developed for Google Earth, is used to save geographic data that includes navigation maps and other driving instructions. Figure (6) shows the live location of an automobile in terms of latitude and longitude, and the engine parameters retrieved by OBD-II: RPM, engine coolant temperature, vehicle speed, percent throttle.

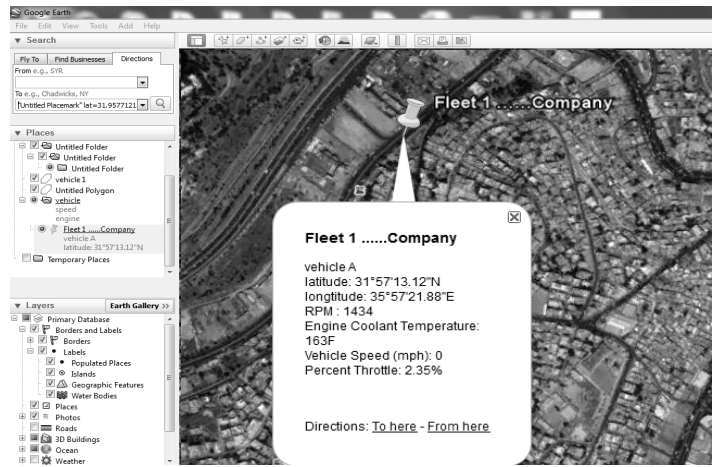

Figure 6. Google Earth Snapshot showing the live location and engine parameters of the tracked automobile.

Furthermore, Google Earth provides the ability to track an object and view the related information at any position as shown in figure (7). The track shows the travel locations of the vehicle form the beginning of the route. All data is saved in a separate excel data sheets.

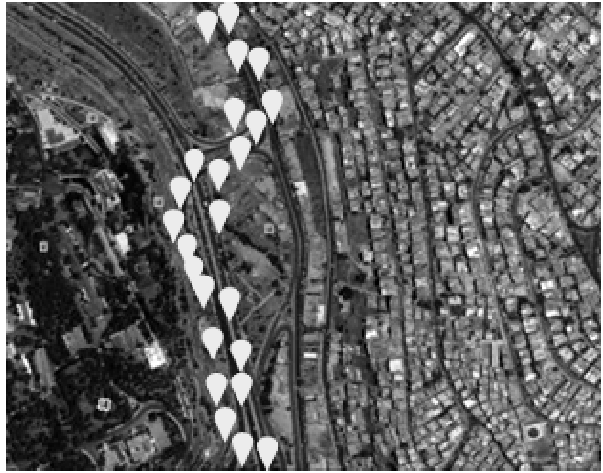

Figure 7. Google Earth Snapshot showing live tracking of targeted automobile.

## 6. Error correction of in-vehicle GPS coordinates using Kalman filtering

The GPS is a satellite-based navigation system. At locations near ground, the satellites signals are reflected by high-rise buildings and other heights, this is known as multipath effect. Therefore, an in-vehicle GPS device may not produce an accurate positioning because of the large delay spreads which cause non line of sight propagation paths of the satellite signals (radio waves). Other factors influencing the accuracy of position determination include: satellite geometry, shifts in the satellite orbits, clock errors of the satellites' clocks, tropospheric and ionospheric effects and calculation errors [15, 16].

To investigate this problem, some researchers proposed mounting 4 GPS antennas onto a vehicle to analyze the correlation of the data from one antenna to the other [17]. This approach will increase the cost and needs more complicated computations. Other researchers studied the effectiveness of differential correction and the influence of well-spaced satellite configurations, where error reduction is done by sending out correction information from fixed earth stations [18].

In this paper, Kalman filter is implemented to reduce GPS errors and thus increase the accuracy of the localization system [19-21]. Our goal is to provide the same precision as Differential GPS (DGPS) systems. The in-vehicle unit transmits the GPS coordinates via GSM module to the reference station, where data is evaluated using Kalman filter to estimate the errors in automobile location [19, 20]. In a GPS measurement system, shown in figure (8), [$S_{xi}$ $S_{yi}$ $S_{zi}$] refers to $i^{th}$ satellite coordinates, [$G_x$ $G_y$ $G_z$] indicates GPS receiver coordinates and $R_i$ represents satellite range as [$S_x$-$G_x$ $S_y$-$G_y$ $S_z$-$G_z$]. Also, pseudorange $PR_i$ is defined as [23]:

$$PR_i = \sqrt{(S_{xi} - G_x)^2 + (S_{yi} - G_y)^2 + (S_{zi} - G_z)^2} + b_u = |R_i| + b_u \quad (1)$$

Where *bu* is receiver clock offset error.

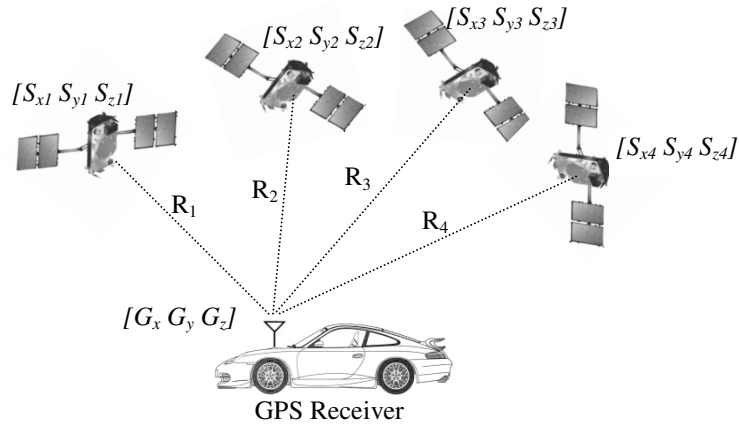

Figure 8. Line-of-sight pseudorange GPS measurements from at least four satellites.

In Kalman filter, a linear and recursive estimator, the states of the system are defined to model the system dynamics. Also, a measurement model is defined to characterize the relationship between the state vector and any measurement. The state vector $x$ of the system at time ($k+1$) are produced by:

$$x_{k+1} = \phi_k x_k + w_k \qquad (2)$$

Where: $\phi_k$ is the state transition matrix. The noise $w_k$ is a white Gaussian noise with zero mean and covariance $Q_k$. To apply Kalman filter in GPS correction procedure, the state vector is defined as:

$$x = [G_x \ G_y \ G_z \ b_u]^T \qquad (3)$$

Where: $[G_x \ G_y \ G_z]$ indicates GPS receiver coordinates, $bu$ is receiver clock offset error. The state transition matrix $\phi_k$ is an identity matrix of 4×4. The process measurement is defined as:

$$z_k = H_k x_k + v_k \qquad (4)$$

Where $H_k$ is the measurement matrix and noise $v_k$ is assumed to be Gaussian with covariance matrix $R_k$. $v_k$ has zero cross-correlation with $w_k$.

The GPS receiver measurement vector for $i^{th}$ satellite includes the pseudorange $PR_i=|R_i|+b_u$ as in equation (1). Linearization of the satellite range $|R_i|$ about estimated GPS receiver coordinates, we find [23]:

$$|R_i| = \sqrt{(S_{xi} - G_x)^2 + (S_{yi} - G_y)^2 + (S_{zi} - G_z)^2} \approx \frac{-(S_{xi}-G_x)\hat{G}_x}{|R_i|} + \frac{-(S_{yi}-G_y)\hat{G}_y}{|R_i|} + \frac{-(S_{zi}-G_z)\hat{G}_z}{|R_i|} \qquad (5)$$

Therefore, the measurement vector is:

$$H_k = \left[\frac{-(S_{xi}-G_x)}{|R_i|} \quad \frac{-(S_{yi}-G_y)}{|R_i|} \quad \frac{-(S_{zi}-G_z)}{|R_i|} \quad 1\right] \qquad (6)$$

The implement of Kalman filter procedure is shown in figure (9). The procedure is initiated by the assumption of $x_0^-$ and $P_0^-$: initial estimate of states and its error covariance respectively. The optimal Kalman gain $K_k$ is utilized to achieve the update estimate of the pseudorange measurements $\hat{x}_k$ and its error covariance $P_k$. The next state $\hat{x}^-_{k+1}$ and error covariance $P^-_{k+1}$ is then calculated based on the current state estimate.

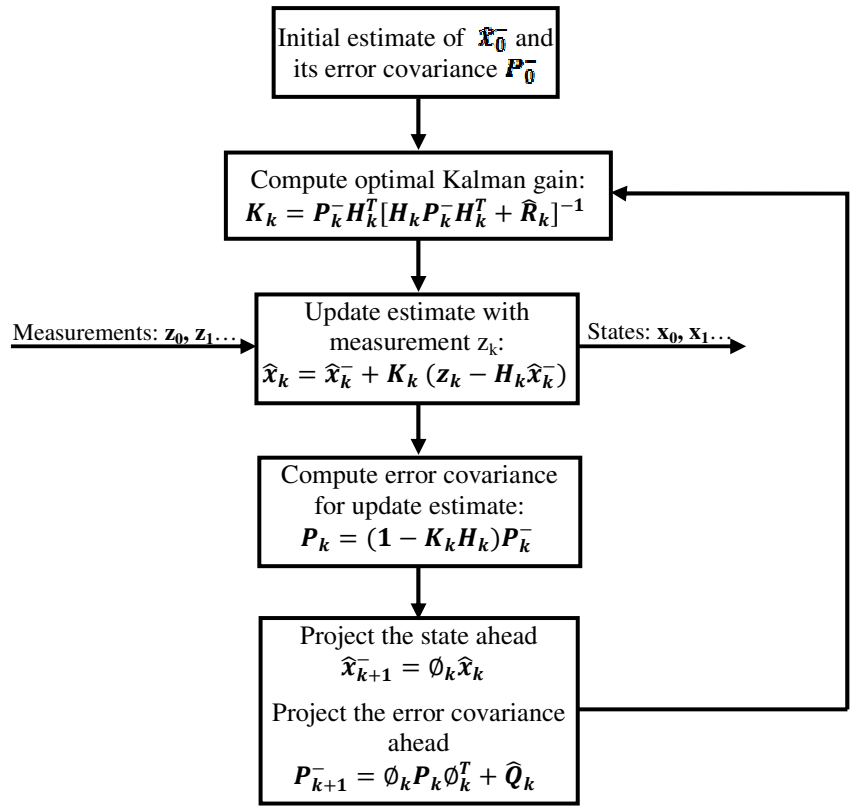

Figure 9. Kalman filter procedure for estimating of GPS receiver coordinates.

The GPS accuracy is measured using *2DRMS* (Twice Distance Root Mean Squared). The computation of *2DRMS* is attained by:

$$2DRMS = 2\left(\sqrt{\sigma_x^2 + \sigma_y^2}\right) \quad (7)$$

Where: $\sigma_x$, $\sigma_y$ are the standard deviations of latitude and longitude respectively of the estimated coordinates by Kalman filter.

The probability represented by *2DRMS* is defined as the typical 95-98% values associated with the probability distribution because the standard deviation of latitude and longitude may not always match.

The results showed *2DRMS* accuracy in the in-vehicle GPS latitude and longitude measurements of around 42.8 meter, figure (10).

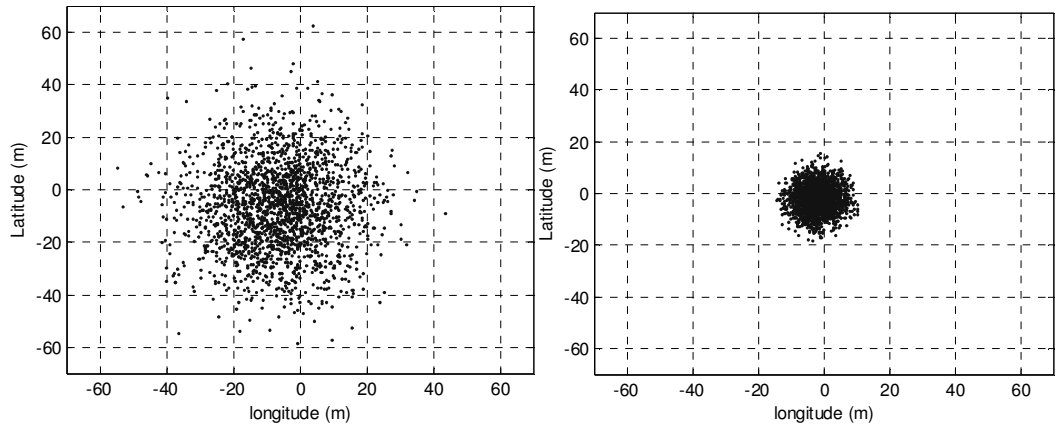

Figure 10. Latitude and longitude measurements for (a) in-vehicle GPS receiver (Latitude standard deviation=13.6 meter, longitude standard deviation=16.5 meter, 2DRMS accuracy= 42.8 meter) and (b) corrected location based on Kalman filter (Latitude standard deviation=5.3 meter, longitude standard deviation=4.3 meter, 2DRMS accuracy= 13.7 meter). The measurements included 2000 data points.

The corrected position is saved by VB to an excel file, which is converted to KML file. The Google Earth shows the information embedded in the KML file. Figure (11) shows the enhancement in the tracking paths for both in-vehicle GPS positioning and corrected GPS readings based on Kalman filter. The resulted 2DRMS accuracy was within the width of the road <14 meter; therefore, the pin icons referring to automobile location were all located at the area of the road as seen in figure (11).

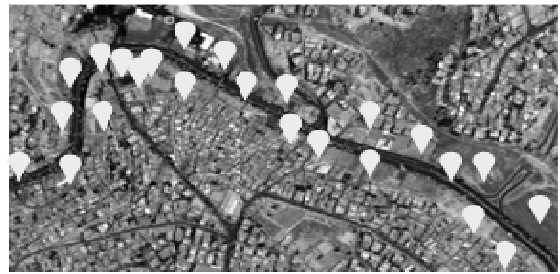

(a)

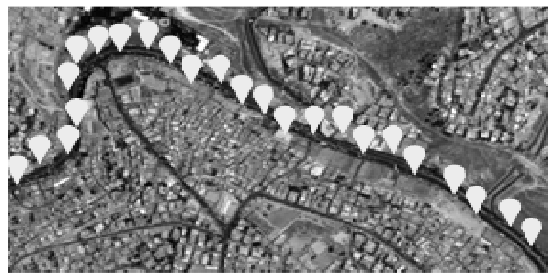

(b)

Figure 11. Google Earth Snapshot showing (a) transmitted GPS position and (b) corrected location of the tracked automobile based on Kalman filter.

Further enhancement of the system could be implemented using map-matching techniques based on the road information to further improve the accuracy of automobile localization.

## 7. Conclusion

In this paper, a real-time automobile tracking system via Google Earth is presented. The system included two main components: a transmitting embedded module to interface in-vehicle GPS and GSM devices in order determine and send automobile location and status information via SMS. The second stationary module is a receiving module to collect and process the transmitted information to a compatible format with Google Earth to remotely monitor the automobile location and status online. The transmitted location of the vehicle has been filtered using Kalman filter to achieve accurate tracking. The *2DRMS* accuracy of estimated vehicle coordinates has been enhanced. The accuracy of filtered coordinates was less than 15 meters compared to about 43 meters for transmitted coordinates received by in-vehicle GPS module.